\documentclass[preprint]{aastex}
\newcommand \kms{km~$\rm{s}^{-1}$}

\newfont{\rten}{cmr10}

\begin{document}


\normalsize

\title{Dust outflows and inner gaps generated by massive planets in debris disks}

\author{Amaya Moro-Mart\'{\i}n\altaffilmark{1} and  Renu Malhotra\altaffilmark{2}}

\email{amaya@as.arizona.edu; renu@lpl.arizona.edu} 

\altaffiltext{1}{Department of Astrophysical Sciences, Princeton University, 
Princeton, NJ 08544, USA}
\altaffiltext{2}{Department of Planetary Sciences, University of Arizona,
1629 E. University Boulevard, Tucson, AZ 85721, USA}

\begin{abstract}
Main sequence stars are commonly surrounded by debris disks, formed by 
cold far-IR-emitting dust that is thought to be continuously replenished 
by a reservoir of undetected dust-producing planetesimals. We have 
investigated the orbital evolution of dust particles in debris disks 
harboring massive planets. Small dust grains are blown out by radiation 
pressure, as is well known; in addition, gravitational scattering by the 
giant planets also creates an outflow of large grains. We describe the 
characteristics of this large-particle outflow in different planetary 
architectures and for different particle sizes. 
In addition, the ejection of particles is responsible for the clearing of 
dust inside the orbit of the planet. We study the efficiency 
of particle ejection and the resulting dust density contrast inside and 
outside the orbit of the planet, as a function of the planet's mass and 
orbital elements and the particle size. We discuss its implications 
for exo-planetary debris disks and for the interpretation of in-situ dust 
detection experiments on space probes traveling in the outer solar system.

\end{abstract}


\section{Introduction}
\label{intro}
Debris disks are disks of dust that surround many main sequence stars.
They were discovered by the IRAS satellite in the 1980's (Aumann et 
al.~\citeyear{auma84}; Gillett~\citeyear{gill86}) and they are preferentially 
detectable at infrared wavelengths, where the dust re-radiates the 
light absorbed from the star.  Stars harboring debris disks 
are too old to have remnants of the primordial disk from which the star itself 
once formed. This is because the dust grain removal processes, such as 
the Poynting-Robertson (P-R) effect and solar wind drag, act on timescales 
much shorter than the age of the star, indicating that such ``infra-red excess
stars'' harbor a reservoir of undetected planetesimals producing dust by 
mutual collisions or by evaporation of comets scattered close to the star 
(Backman \& Paresce~\citeyear{back93}). 
The spectroscopy of systems like $\beta$-Pictoris supports this interpretation 
(e.g. Knacke et al.~\citeyear{knac93}; Pantin, Lagage \& Artymowicz~\citeyear{pant97}).
It seems clear, therefore, that planetesimals are present in these systems. 
But what about massive planets? High-resolution images of some of these debris 
disks have revealed the presence of density structure (see Koerner~\citeyear{koer01} 
for a review) and dynamical models have shown that planets can sculpt the dust disks, 
creating gaps, arcs, rings, warps and clumps of dust (e.g. Roques et al.~\citeyear{roqu94};
Liou \& Zook~\citeyear{liou99}; Mouillet et al.~\citeyear{moui97}; 
Wyatt et al.~\citeyear{wyat99}; Moro-Mart\'{\i}n \& Malhotra~\citeyear{ama02};
Kuchner \& Holman~\citeyear{kuch03}). The combination of both, the very high 
resolution imaging at long wavelengths and theoretical dynamical models can 
provide interpretation of the disks' structure in terms of planetary architectures. 
This approach has been used in the interpretation of high resolution 
millimeter interferometry observations of the Vega system (Wilner et al.~\citeyear{wiln02}) 
and of the submillimeter images of the $\epsilon$ Eridani system 
(Ozernoy et al.~\citeyear{ozer00}; Quillen \& Thorndike~\citeyear{quil02}).
Recent observations with the $\it{Spitzer}$ MIPS instrument have confirmed that 
out of 26 FGK field stars known to have planets by radial velocity studies, 6 
show 70$\mu$m excess at 3-$\sigma$ confidence level, implying the presence of 
cool material ($<$100 K) located beyond 10 AU (Beichman et al.~\citeyear{beic05}). 
These stars, with a median age of 4 Gyr, are the first to be identified as having 
both well-confirmed planetary systems and well-confirmed IR excesses (Beichman 
et al.~\citeyear{beic05}). In addition, the first results from the $\it{Spitzer}$ 
$\it{FEPS}$ Legacy project indicate that inner 
gaps\footnote{In this paper a ``gap'' is an inner depletion zone in the dust disk
interior to the planet's orbit, not an annular depletion zone around the planet's orbit.}
appear to be common in cold Kuiper Belt-like disks (Kim et al.~\citeyear{kim05}). 
These disks show excesses at 70 $\mu$m but not at 24 $\mu$m, indicating again the 
presence of cool dust ($<$100 K) located beyond 10 AU. The lack of 24 $\mu$m 
emission yields an upper limit to the amount of warm dust inside 10 AU; this 
upper limit is 10$^{-3}$ to 10$^{-2}$ times the lower limit for the mass in 
the corresponding cold disk. Because the lifetime of the dust particles due 
to P-R drag is of the order of 1 Myr, it is expected that the density contrast 
would be erased on this timescale. Kim et al. (\citeyear{kim05}) suggest that a 
possible explanation for these inner gaps is that one or more massive planets 
are dynamically depleting, via gravitational scattering, dust particles generated 
by an outer belt of planetesimals. All these observations are providing increasing 
evidence that debris disks and massive planets co-exist around many sun-like stars.

In this paper we report some new results based on numerical modeling regarding the 
depletion of large dust particles in debris disks by the gravitational perturbations 
of massive planets. The numerical models used to carry out this study are briefly 
described in $\S$2. The ejected particles form an ``outflow'' whose properties 
(angular confinement, velocity and efficiency of ejection) are characterized in 
$\S$3.1 as a function of the planet's mass and orbital elements, and the particle 
size. The high efficiency of ejection, together with the possible high frequency 
of debris disks harboring massive planets, suggest that these outflows may be a 
common phenomenon, whose implications are described in $\S$3.2. The ejection of 
particles is also responsible for the depletion of dust interior to the orbit of 
the planet, creating a density contrast that can be measured directly in 
spatially resolved images or indirectly through the modeling of the spectral 
energy distribution (SED) of the debris disk. To aid in the interpretation of
such observations, in $\S$4 we study the density contrast inside and outside 
the orbit of the planet, as a function of the planet's mass and orbital elements 
and the particle size. Finally, $\S$5 summarizes our results.

\section{The numerical models}
We numerically solve the equations of motion of dust particles 
generated in a debris disk, analogous to the solar system's Kuiper Belt.
We use a modified version called SKEEL of the multiple time 
step symplectic method SyMBA (Duncan, Levison \& Lee~\citeyear{dunc98}; 
Moro-Mart\'{\i}n \& Malhotra~\citeyear{ama02}).
Our models include the combined effects of solar gravity, solar radiation 
pressure, the P-R effect and solar wind drag, and the gravitational forces 
of planets. We model the solar system with 7 planets (excluding Mercury 
and Pluto, and including the mutual perturbations of the planets), 
and we model hypothetical extra-solar planetary systems with single planets 
of different masses, semimajor axes and eccentricities (see Tables 1, 2 and 3
for a complete list of models). 
For some of these systems, the parent bodies of the dust particles 
are assumed to be distributed in orbits with semimajor axis between 35 and 
50 AU, eccentricities such that the perihelion distances are between 35 and 
50 AU, and inclinations between 0 and 17$^\circ$, in approximate accord with 
current estimates of the orbital distribution of the classical Kuiper Belt 
(Malhotra et al.~\citeyear{malh00}; Brown~\citeyear{brow01}).
For other systems, the dust-producing planetesimals are randomly distributed 
in a thinner disk with $\it{a}$=35--50 AU, $\it{e}$=0--0.05 and $\it{i}$=0--0.05 radians.
In all our models, the initial values of mean anomaly 
(M), longitude of ascending node ($\Omega$) and  argument of perihelion 
($\omega$) were randomly distributed between 0 and 2$\pi$. We run models for
different particle sizes, referred to in terms of their $\beta$ value, 
which is the dimensionless ratio of the radiation pressure force and the 
gravitational force. For spherical grains, 
\begin{equation}
{\beta = (3L_{*}/16 \pi \it{GM_{*}c})(Q_{pr}/\rho~s),}
\end{equation}
where L$_{*}$ and M$_{*}$ are the stellar luminosity and mass; for a solar-type star, 
$\beta$=5.7 $\times$ 10$^{-5}$ Q$_{pr}$/$\rho\cdot$s, 
where $\rho$ and s are the density and radius of the grain in cgs units 
(Burns, Lamy \& Soter~\citeyear{burn79}). Q$_{pr}$ is the radiation 
pressure coefficient, a function of the physical properties of the grain 
and the wavelength of
the incoming radiation; the value we use is an average, integrated over the 
solar spectrum. [For the correspondence between $\beta$ and the particle size see
Fig. 5 in Moro-Mart\'{\i}n, Wolf \& Malhotra (\citeyear{ama05}).]
The sinks of dust included in our numerical simulations are (1) 
ejection into unbound orbits, (2) accretion into the planets, and 
(3) orbital decay to less than 0.5 AU heliocentric distance (0.1 AU 
for the models with a single planet located at 1 AU).  
A detailed description of the numerical algorithm used to integrate the equations 
of motion is given in Moro-Mart\'{\i}n \& Malhotra (\citeyear{ama02}).

\section{Dust outflows from debris disks}
\label{dustoutflow}
Radiation pressure arises from the interception by the dust particles of the 
momentum carried by the incident stellar photons; it makes the orbits of the 
dust particles change immediately upon release from their parent bodies
(i.e., the meter-to-kilometer size dust-producing planetesimals).
For parent bodies in circular orbits, small grains with $\beta >$ 0.5 
are forced into hyperbolic orbits as soon as they are released. If the 
parent bodies' orbits are eccentric, ejection occurs for 
$\beta >$ 0.5(1$\mp${\it e}) for a particle released at perihelion or 
aphelion, respectively. In the solar system these particles are known as 
$\beta$-meteoroids (Zook \& Berg~\citeyear{zook75}). 
These small dust particles leave the system in a ``disk wind'', whose 
angular extent is determined by the inclinations of the 
parent bodies; this is because radiation pressure is a radial force which 
does not change the inclinations of the dust particles after their release. 

Grains larger than the ``blow-out'' size, on the other hand, remain on bound 
orbits upon release, and their orbital evolution is the subject of our study.
Their dynamical evolution is affected by the P-R effect, which tends to circularize 
and shrink their orbits, forcing these particles to slowly drift in toward 
the central star (Burns, Lamy \& Soter~\citeyear{burn79}).
If no planets were present, the final fate of these dust particles would
be to drift all the way into the star until they sublimate. Other removal 
processes may include mutual grain collisions and collisions with 
interstellar grains, which may comminute the grains to sizes small enough to 
be blown away by radiation pressure. [The studies reported here do not include 
collisional effects; for an estimate of the limitations of our models we refer 
to Moro-Mart\'{\i}n \& Malhotra (\citeyear{ama02} and \citeyear{ama03}).]
When planets are present the story changes: (a) the trapping of particles in 
mean motion resonances (MMRs) with the planets causes an accumulation of 
particles at resonant semimajor axes; and (b) sufficiently massive 
planets can scatter and eject dust particles out of the planetary system.
In the case of dust produced in the Kuiper Belt in our 
solar system, about 80--90\% of the dust grains are ejected 
by close encounters with the giant planets (mainly Jupiter and Saturn), 
a few percent accrete onto the 
planets, and the remaining 10--20\% drift all the way into the Sun 
(Liou, Zook \& Dermott~\citeyear{liou96}; Moro-Mart\'{\i}n \& 
Malhotra~\citeyear{ama03}; see also Table 1).
Thus, in addition to the afore-mentioned $\beta$-meteoroids, an outflow of larger 
particles produced by gravitational scattering from planets also exists.

\subsection{Dependence on planetary architecture and particle size}
We have explored the characteristics of the large particle outflow and its 
dependence on planetary architecture and particle size. For the solar 
system architecture, it is known 
that the majority of KB dust particles are ejected by Jupiter and Saturn
(Liou, Zook \& Dermott~\citeyear{liou96}; Moro-Mart\'{\i}n \& Malhotra~\citeyear{ama03}). 
Motivated by this, we have modeled hypothetical planetary systems consisting of a 
single planet and a KB-like dust source. These models explore a 
range of planetary masses (M$_p$/M$_{Jup}$=0.03, 0.1, 0.3, 1, 3, and 10), orbital 
semimajor axis ($\it{a}$=1, 5.2, 10, 20 and 30 AU), and eccentricities ($\it{e}$=0, 
0.1, 0.2, 0.3, 0.4, 0.5)  (see Tables 1, 2 and 3). 

Fig.~1 shows examples of the escaping\footnote{Our definition of ``escaping'' 
is that the particles reach a distance 1000 AU from the star (see Fig. 3 to 6); 
at that point, we stop integrating their orbits. This is not quite equivalent to the precise 
criterion for ejection, which would be that a particle velocity exceed the 
escape velocity. However, our numerical studies find that the particles that 
reach 1000 AU, 30-60\%  (depending on their $\beta$) are in hyperbolic orbits,
and more than 90\% have orbital eccentricity $e>$0.98. 
This means that even though some of the particles are still bound by 
the time they reach 1000 AU, it is very likely that they will also be 
set on hyperbolic orbits within a few orbits, either by subsequent 
scattering from the planets or due to small additional perturbations 
not included in our models.} 
particle trajectories for the solar system case, projected in the ecliptic 
plane (XY; left panel), and in the RZ plane (right panel; where R is the 
in-plane heliocentric distance and Z is the off-plane out-of-ecliptic distance).
These examples are of particles that reach at least 1000 AU and had their last 
encounter with Jupiter. We see that Jupiter creates a fan-like outflow, 
mainly confined to the ecliptic, where the trajectories are in the 
counterclockwise (prograde) direction. The distributions of eccentricity, 
inclination and perihelion of these Jupiter-ejected particles are presented 
in Fig. 2. The histograms show that all the particles are either in or very close 
to hyperbolic orbits; that the scattering rarely changes the inclination of the 
particles by more than 15 degrees (see also column 9 in Table 1); and that 
few of the ejected particles leave on orbits of perihelion interior to Jupiter's 
orbit. 

For the single planet models, Fig. 3--6 show the velocities of the 
escaping particles at 1000 AU projected in the XY (ecliptic) plane (left) 
and in the XZ plane (right). At large heliocentric distances the outflow 
is radial and symmetric, except when the planet is in an eccentric orbit 
(Fig. 6); the projection in the XZ plane shows that it is largely confined 
to the ecliptic for Jupiter-mass planets (or smaller), and becomes less
confined as the planet mass increases. The angular confinement to the disk 
can also be seen in Fig. 7 and 8, in the distribution of orbital inclination 
for the ejected particles, and in column 9 of Tables 1, 2 and 3. This angular 
confinement is not obvious a priori because the ejection of the particles 
is due to gravitational scattering, a process that does not necessarily 
preserve the inclination of the orbits.

In Tables 1--3, we give a list of the single-planet and multiple-planet models 
that we have simulated (a total of 126 models). Also included in these tables 
are the statistical results for the fates of the dust particles in each model. 
We have performed simulations for several solar system models with the same or 
similar initial conditions of the dust parent bodies and the results indicate 
that a conservative estimate of the uncertainty in n$_{1000}$, owing to the 
chaotic dynamics of dust orbital evolution, is $\sim$10\% of the initial number 
of particles. 

Fig. 9 and 10 show the percentage of particles that are gravitationally scattered 
out from the system, and the velocity at infinity of the ejected particles, as 
a function of the planet's mass, semimajor axis and eccentricity, and the particle 
size. We find the following dependencies (the parentheses show the values explored 
by our models).
\begin{itemize}
\item {\it Particle sizes} 
($\beta$=0.00156, 0.00312, 0.00625, 0.0125, 0.025, 0.044, 0.1, 0.2 and 0.4):
It is expected that gravitational scattering is dependent to some extent on
the particle size as smaller particles (larger $\beta$) migrate past the 
planet faster, therefore decreasing their probability of ejection. 
The top panel of Fig. 9 shows that:
(1) For a 1M$_{Jup}$ planet, the efficiency of ejection decreases as $\beta$ 
increases, reaching a minimum at $\beta$$\sim$0.1--0.2 and increasing thereafter. 
As mentioned above, the decrease in efficiency is expected because the particle
P-R drift velocity is larger for larger $\beta$. The increase in efficiency for 
even larger $\beta$ is probably due to the fact that radiation pressure is starting 
to contribute to the ejection of the particles. 
(2) The effect described above is more significant for close-in planets (1 AU), 
i.e. when the particle is deeper in the potential well of the star. 
(3) Planets $>$3M$_{Jup}$ in circular orbits between 1 AU and 30 AU eject
$>$80\% of the particles that go past, independently of the particle size.
In addition, from the top panel of Fig. 10, we see that
there is an increase in $\it\bar{v}_\infty$ as the particle size 
decreases ($\beta$ increases), which is more pronounced when the perturbing
planet is closer to the star.  The distributions of particle inclinations in
Fig. 2 and 7 show that the angular confinement of the ejected particles is similar 
for all particle sizes. This is not surprising because the inclination perturbation 
in gravitational scattering is independent of particle size, as particle masses 
are more than 30 orders of magnitude smaller than the masses of the planets.

\item {\it Planet semimajor axis} (1, 5.2, 10, 20 and 30 AU): We find that the 
average dust outflow velocity is larger in the presence of close-in planets 
than more distant planets of the same mass (see top panels of Fig. 10).
This trend is clearly seen in the left panel of Fig. 11; the slope of the line 
corresponds to approximately $\it\bar{v}_\infty$$\varpropto$$\it{a}$$_{pl}^{-0.5}$, 
and is consistent with an analytical calculation by Murray, Weingartner \& 
Capobianco (\citeyear{murr03}).

\item {\it Planet mass} (0.03, 0.1, 0.3, 1, 3, and 10M$_{Jup}$): 
The right panel of Fig. 11 shows only a weak dependence on the mass of the 
planet of the average particle ejection velocity; this is somewhat in contrast
with the theoretical prediction, $\it{v}_\infty\varpropto$M$_{pl}^{1/4}$
(Murray, Weingartner \& Capobianco~\citeyear{murr03}). 
The magnitude of the ejection velocity, $\sim$3 \kms, in the Jupiter-mass
single-planet models (blue line in the top left panel of Fig.~10)
is higher than the numerical result in Murray, Weingartner \& 
Capobianco (\citeyear{murr03}), but  agrees better with their analytical 
estimate.  Their analysis, however, assumes that the particle ejection takes place 
after only a single planetary encounter, whereas our simulations show that 
typically ejections occur after many planetary encounters. (In our simulations, 
we track the planetary encounters of dust particles within 3.5 Hill-radius 
distance from each planet. The number of such encounters that ejected particles 
suffer is on the order of 10--10$^4$, with the lower range being more typical 
in models with more massive planets, 3--10 M$_{Jup}$). 
In addition to this complexity, 
it is important to remember that the effect of the planet's orbital 
elements and mass on the outflow parameters (velocity and confinement 
to the plane) is not only direct, via the close encounters, but also 
indirect, as the particles encounter the planet with a history of 
evolution in the MMRs that can change the initial orbital elements 
of the particles and therefore affect their subsequent dynamical 
evolution. As an example, the eccentricity distributions of the 
soon-to-be-ejected particles near the planet show that for the 1 
and 3 M$_{Jup}$ models, $\it{e}\sim$0.4--0.5, but for 10 M$_{Jup}$, 
$\it{e}<$0.2. 

The distribution of inclinations in Fig. 7 shows that for a planet at 1 
and 5.2 AU, the angular confinement of the outflow to the disk is affected by 
the planet's mass; the more massive the planet the less confinement the outflow 
has. However, the parameter that is most strongly dependent on the planet's 
mass is the number of ejected particles.
The bottom left panel of Fig. 9 shows that there is a sharp increase in 
ejection efficiency when the planet mass increases from 0.3 M$_{Jup}$ to 
1 M$_{Jup}$: planets $\lesssim$0.1 M$_{Jup}$ do not eject a significant
number of particles, whereas planets  $>$3 M$_{Jup}$ eject $>$90\% 
if located between 1--30 AU.
A 1 M$_{Jup}$ planet at 5--30 AU ejects about 80\% of the particles, 
and about 60\% if located at 1 AU.

\item {\it Planet eccentricity} (0, 0.1, 0.2, 0.3, 0.4 and 0.5): 
Large planet eccentricities create an asymmetric outflow oriented along 
the major axis of the planet's orbit. The number of particles ejected in 
the apoastron direction exceeds that in the periastron direction by a 
factor of $\sim$5 for e=0.5 (see Fig. 6). The asymmetry is due to the fact that the 
planet spends more time near apoastron and therefore the probability of encounter 
with a dust particle is higher near apoastron. Fig. 8 and 10 show that the 
inclinations and the average velocity of the ejected particles at infinity 
are not affected by the planet's eccentricity. The efficiency
of ejection, however, decreases significantly as the planet's eccentricity 
increases: for a 1 M$_{Jup}$ planet at 5 AU it decreases from 
$\sim$80\% to $\sim$30\% when the planet eccentricity is 
increased from 0 to 0.5 (see bottom right panel in Fig. 9). 
It is of interest to note that many of the known exo-planets to date have 
large orbital eccentricities (Marcy et al.~\citeyear{marc03}); our
models predict that the large particle outflow will be asymmetric 
in these cases.

\item {\it Comparison with Solar System}: 
The single-planet analog of the solar system (i.e. only Jupiter in a circular 
orbit at 5.2 AU) produces a somewhat higher velocity outflow compared with the 
actual multi-planet solar system. This is mainly due to the effect of Saturn in 
our solar system: having a larger semimajor axis, Saturn intercepts a fraction 
of the KB dust grains as they evolve inward due to the P-R drag and ejects them 
at a somewhat lower velocity, thus depressing the mean velocity of the outflow.
\end{itemize}

\subsection{Implications of dust particle outflows}
There are several significant implications of this large-particle outflow.
\subsubsection{Exo-planetary debris disks and planet formation environment}
Stellar surveys show that at least 15\% of A-K main sequence stars 
are surrounded by debris disks, and that the far-infrared 
excess decreases with stellar age, 
dropping from about 50\% to about 15\% after approximately 500 Myr. 
But these samples are sensitivity-limited, and therefore the 
occurrence of debris disks could be higher (Lagrange, Backman \& 
Artymowicz~\citeyear{lagr00} and references therein). 
Stellar radial velocity surveys indicate that about 7\% of the FGK main 
sequence stars have a Saturn or Jupiter-mass planet within 3 AU 
(Marcy~\citeyear{marc03}). Even though the correlation between
the presence of planets and debris disks is not known yet, 
our studies suggest these large-particle dust outflows
may be a common phenomena in planetary systems that harbor debris disks. 
This is of interest because:\\
(a) These large-particle dust outflows {\it may contribute significantly or even 
dominate the clearing of circumstellar debris in planetary systems}.
Hitherto, the main processes that have been considered for such clearing are 
stellar winds, radiation pressure, sublimation, and collisions. The latter reduce 
the size of the dust particles until they are small enough to be blown away by 
radiation pressure. However, as our models indicate, 
gravitational scattering by giant planets following orbital decay by P-R drag 
is also significant, and in some cases may be a dominant process, ejecting 50--90\% 
of the dust grain population.\\
(b) These outflows should be added to the list of processes that 
{\it link the interplanetary environment to the galactic environment 
of a star}. Planetary systems are prime sites for large particle formation. 
As such, they can contaminate the immediate vicinity of star-forming regions 
through this large particle outflow, and thus affect the particle size 
distribution of their local ISM. It is likely, therefore, that large particle 
outflows from extra-solar planetary systems may be a source of the large 
interstellar particles that have been detected in the interplanetary medium.

The presence of an outflow in an exo-planetary system and its detectability 
will strongly depend on the orbital characteristics of the planet and the 
orientation of the system. For face-on systems, the expected surface brightness 
of the dust outflow will be very low, making it very hard to detect astronomically 
as a radial extension of the debris disk. Additionally, the lack of velocity 
information from usual infrared measurements will not allow to distinguish 
between an outflow and a bound disk. 

The face-on optical depth of a disk composed of grains of 
radius $\it{a}$ and observed at frequency $\nu$ is given by 
(Backman \& Paresce~\citeyear{back93}): 
$\tau$$_\bot$($\it{r}$,$\nu$)= $\sigma$($\it{r}$)($\xi$$\it{a}$$\nu$/c)$^q$; 
where $\sigma$($\it{r}$) cm$^2$/cm$^2$ is the face-on fractional 
geometric surface density; it is equal to the surface density $\it{n}$($\it{r}$), 
multiplied by the geometric cross section of the grain, 
$\sigma$($\it{r}$)=$\it{n}$($\it{r}$)$\pi\it{a}^2$. 
$\xi$ is the ratio between the critical wavelength $\lambda$$_0$ 
up to which the grain absorbs and emits radiation efficiently)
and the grain radius $\it{a}$, and depends on the grain properties
(e.g. $\xi$$\equiv$$\lambda$$_0$/$\it{a}$$\sim$2$\pi$,
1/2$\pi$ and 1, for strongly, weakly and moderately absorbing 
materials; we will use $\xi$$\sim$1).
$\it{q}$ is the power law index of the emissive efficiency
$\epsilon$, such that for $\lambda$$<$$\lambda$$_0$, $\epsilon$$\sim$1,
but for longer wavelengths the emissive efficiency decreases as 
$\epsilon$=$\epsilon$$_0$($\lambda$$_0$/$\lambda$)$^q$; for the 
intermediate size regime, where $\it{a}$ is larger than $\lambda$$_{peak}$
of the incoming radiation (absorbs efficiently) but smaller than 
$\lambda$$_{peak}$ of the grain thermal emission (emits inefficiently),
$\it{q}$=1. And c is the velocity of light.

We can estimate the surface density $\it{n}$($\it{r}$) (cm$^{-2}$) 
at a distance $\it{r}$ from the central star from mass 
conservation by equating the mass that is produced in time dt, 
dN=$\it{dpr}$$\it{f}$$_{ej}$dt, with the mass that crosses 
the annulus of radius $\it{r}$ in time dt, 
dN=$\it{n}$($\it{r}$)2$\pi$$\it{r}$$\it{v}$dt. 
$\it{dpr}$ is the dust production rate in particles per second; 
$\it{f}$$_{ej}$ is the fraction of particles that are ejected
(our numerical studies find $\it{f}$$_{ej}$$\sim$50--90\%); and $\it{v}$ is the
velocity of the particles at distance $\it{r}$, for large distances 
we will take
$\it{v}$$\approx$$\it{v}$$_{esc}$=(2$\it{G}$M$_{\sun}$/$\it{r}$)$^{1/2}$.
Solving for $\it{n}$($\it{r}$) and substituting into $\sigma$($\it{r}$),
\begin{equation}
{\tau_{\bot}^{outflow}(r,\nu)=\sigma (r)({\xi a \nu \over c})^q=
{\it{dpr}\it{f}_{ej}\over 2\pi\it{r}(2GM_{\sun}/r)^{1/2}}\pi\it{a}^2 ({\xi a \nu \over c})^q.}
\end{equation}
We can estimate the optical depth of the solar system's outflow using
the KB dust production rates derived by Landgraf et al. (\citeyear{land02}), 
which are based on {\it Pioneer 10} and {\it 11} measurements and   
for the Kuiper Belt gives $\it{dpr}$$\sim$2$\times$10$^{14}$ particles/s
(for particles between 0.01 and 6 mm). Because the size  
distribution is very steep, one can assume that most of 
the detections are caused by particles just above
the detection threshold, i.e. particles with $\it{a}$$\approx$5 $\micron$.
For this particle size, $\beta$$\approx$0.05 and 
$\it{f}$$_{ej}$$\approx$0.8, and the optical depth 
at 60$\micron$ ($\nu$=5$\times$10$^{12}$Hz) will then be
$\tau$$_\bot$$^{outflow}$($\it{r}$,$\nu$)=
2.6$\times$10$^{-14}$/$\it{r}$$^{1/2}$ (where $\it{r}$ is in AU).
We can compare this to the optical depth of the Kuiper Belt (bound) disk. 
From Fig. 11 in Moro-Mart\'{\i}n \& Malhotra (\citeyear{ama02}) we can get 
the surface density that corresponds to a fictitious dust
production rate of 100 particles per 1000 years, 
$\it{n}$$\approx$300 particles/AU$^2$. Scaling up
this density to account for the dust production rate found by  
Landgraf et al. (\citeyear{land02}), we find that
$\it{n}$$\approx$8.4$\times$10$^{-2}$ particles/cm$^2$, 
$\sigma$$\approx$6.6$\times$10$^{-8}$, so that
$\tau$$_\bot$$^{disk}$$\approx$5.5$\times$10$^{-9}$. 
For the solar system, the ratio of the two optical depths is then
$\sim$10$^{-6}$.
Other models for the Kuiper Belt dust disk
give $\sigma$$\approx$10$^{-6}$ (15 times larger than our value; 
Backman, Dasgupta \& Stencel~\citeyear{back95}). 
It is estimated that for a system at 30 pc, the 70 $\micron$ $\it{MIPS}$ 
array in $\it{Spitzer}$ will be able to detect a disk with 
$\sigma$$\approx$3$\times$10$^{-6}$ (D. Backman, private communication).
This means that in order to see the Kuiper Belt dust disk the dust production 
rate will need to be increased by a factor of $\sim$3 in Backman's models, or a factor of 
$\sim$45 in our models (using Landgraf's $\it{dpr}$). But in order to see the 
outflow it will need to be increased by a 
factor of $\sim$6$\times$10$^6$ (Backman's) or 9$\times$10$^7$
(ours). In any case, this increase will make the bound disk 
be optically thick.  In other words, for an optically thin debris disk 
(where our dynamical models are valid), this outflow is very unlikely to be
detected.  For younger and more massive edge-on systems, after the giant 
planets have already formed, it may be possible to detect the outflow out 
of the plane. In this geometry, the signature of the off-plane outflow will 
be clearer against the fainter background.
However, our dynamical models are not valid in this high-density regime where collisional 
effects dominate over P-R drag. 
It is possible that such an outflow may have already been detected with 
the Advanced Meteor Orbit Radar, which senses plasma signatures 
produced by extra-terrestrial dust particles ablating in the Earth's 
atmosphere: Taylor, Baggaley \& Steel (\citeyear{tayl96}) and Baggaley 
(\citeyear{bagg00}) claim that the main discrete source seems to coincide 
in direction with $\beta$ Pictoris. 

\subsubsection{Interpretation of in-situ dust detections made by space probes}
Recent Ulysses and Galileo dust experiments have led to the surprising 
discovery of interstellar grains sweeping through the solar system deep 
within the heliosphere (Grun, Zook \& Baguhl~\citeyear{grun93}).
Previously, interstellar grains could only be studied by extinction 
and polarization measurements of optical starlight, not sensitive to 
grains larger than 0.3 microns because of their small contribution to 
the optical cross section, and by infrared emission. These in-situ 
detections allowed us for the first time to study the mass distribution 
of interstellar grains within the heliosphere, leading to the surprising
discovery of a population of large particles ($> 10^{-16}$ kg, Grun et 
al.~\citeyear{grun94}) that 
are 30 times more massive than the interstellar grains that cause stellar
extinction. This finding implies that more mass is locked up in large 
grains locally than has been estimated from the astronomical measurements.
The gas-to-dust ratio derived from astronomical measurements
(400--600) is found to be much larger than the value of $\sim$100 derived from 
the in-situ detections, implying that the local interstellar cloud exceeds 
cosmic abundances (Frisch et al.~\citeyear{fris99}).
These very important results rely critically on the correct identification 
of the origin of the dust grains. This identification 
is based on a geometrical argument: the direction
the grains are coming from, with interstellar grains coinciding 
with the flow of neutral helium through the 
solar system; and a dynamical argument: the impact velocity
and the expectation that only interstellar grains are on unbound hyperbolic
orbits (Grun, Zook \& Baguhl~\citeyear{grun93}).
Under the current understanding, the sources of meteoroids
in interplanetary space and their orbital properties are assumed as follows:  
Asteroids: low eccentricity and inclination; Comets:  
high eccentricity and inclination; Kuiper Belt: low eccentricity 
and inclination; and Interstellar: hyperbolic, and aligned with the 
direction of flow of the interstellar gas. However, we have shown 
in this paper that $\sim$80--90\% of 
large Kuiper Belt grains ($\beta<0.5$) are gravitationally scattered outward 
by Jupiter and Saturn into hyperbolic orbits; therefore there is the potential of 
misinterpreting these escaping interplanetary particles as interstellar. 
In addition, other sources exist such as comets, Asteroid
Belt and Trojan asteroids. Due to radiation pressure, some of the 
dust particles released at those locations will be set on Jupiter 
crossing orbits, so in principle close encounters with Jupiter could take place
resulting on hyperbolic orbits. In the future, we plan to study whether or not
these particles may have been detected by Ulysses and Galileo.
For the analysis of future in-situ dust detections in the outer solar system, such 
as with the {\it Cassini} Cosmic Dust Analyzer and the {\it Interstellar Probe}, 
it will be important to keep in mind the existence of the large-particle
outflow of solar system dust to correctly identify the origin of the massive fast moving 
particles, whether interplanetary or interstellar. It has been recently
announced that the analysis of the ion charge signals in the 
{\it Cassini} dust detector, together with geometric and kinematic
considerations, have led to the identification of an interstellar flux
at 0.8 AU that is in agreement with the flux measured by Ulysses at 3 AU at 
the same time (Altobelli et al \citeyear{alto03}).  
But any dust detections by Cassini outside Jupiter's orbit have not yet been reported.

\section{On how debris disks with inner gaps signal the presence of massive planets}
\label{gaps}
Recent GTO and $\it{FEPS}$ observations with the $\it{Spitzer}$ MIPS 
instrument suggest that debris disks and giant planets co-exist and that
inner gaps appear to be common in cold Kuiper Belt-like disks (Beichman 
et al.~\citeyear{beic05} and Kim et al.~\citeyear{kim05}). In view of these 
observations, it is interesting to study the efficiency of particle ejection 
($\S$3.1) and the resulting dust density contrast inside and outside the orbit 
of the planet, as a function of the planet's mass and orbital elements and the 
particle size. It is important to keep in mind, however, that the modeling presented 
here does not consider the effect of particle collisions, which together with P-R 
drag could also be responsible for the opening of an inner gap in the
dust disk (Wyatt~\citeyear{wyat05}).  

If the particles were drifting inward at a constant rate, as set by 
P-R drag, the ratio n$_{in}$/n$_{1000}$ (from Tables 1, 2 and 3) would 
directly give us an estimate of the density contrast inside and outside 
the inner boundary of the disk. However, the trapping of particles in MMRs
with the planet halts the P-R drift, increasing the number density of 
particles in that region. The density contrast, therefore, can only be 
estimated using the radial density profiles that result from the numerical 
simulations. Fig. 12 shows some of these profiles for a representative set 
of models. These results, keeping in mind the uncertainties due to the fact that we 
are modeling the dynamical evolution of a small number of test particles 
(N$\sim$100), can help us estimate what planet masses and semimajor axes 
may be responsible for the inner gaps that are inferred indirectly from the 
disks' SEDs, or in few cases, that are seen directly in spatially resolved 
images. For planets located at 1--30 AU with masses of 1--10M$_{Jup}$, 
the ratio between the density outside and inside the orbit of the planet is 
$\gtrsim$40, whereas for planet masses of 0.03--0.3M$_{Jup}$, this ratio is 
in the range 3--10. The models show that the radius of the inner depleted
region, r$_{gap}$, depends on the mass and the eccentricity of the planet. 
In Table 4 we show that for the models with planets in circular orbits, 
r$_{gap}$$\sim$0.8$\times$a$_{pl}$ for 1--3M$_{Jup}$ and $\sim$1.2$\times$a$_{pl}$ 
for 10M$_{Jup}$. The three bottom planels of Fig. 12 show that for planets 
with eccentricities in the range 0.3--0.5, the surface density decreases more
smoothly and consequently the dust disk would not present a sharp inner edge.

\section{Conclusions}
\label{conclusions}
When a massive planet is located interior to a belt of dust-producing 
planetesimals, dynamical models have shown that as the dust particles 
drift inward due to P-R drag, they get trapped in MMRs with the planet, and
this well-known effect can sculpt the dust disk creating rings, warps and azimuthal 
asymmetries. In addition to the trapping in MMRs, gravitational scattering
with the planet is responsible for the depletion of dust inside the orbit
of the planet. Although this is also a well known effect, to our knowledge 
it has not been studied in detail in the past. In this paper we have shown
that the ejected dust particles form an ``outflow'', whose angular 
confinement, velocity and symmetry depend on the planet's mass and orbital 
elements, as well as the particle size. The high efficiency of ejection
(for planet masses $\gtrsim$1M$_{Jup}$), together with the possible high frequency 
of debris disks harboring massive 
planets, suggest that these outflows may be a common phenomenon. If this
is the case, they may contribute significantly or even dominate the clearing 
of circumstellar debris in planetary systems, enriching the immediate vicinity 
of star-forming regions with large dust particles and affecting therefore the particle 
size distribution of their local ISM. In addition, we have seen how the ejection 
of particles is responsible for the clearing of dust inside the orbit of 
the planet, creating a density contrast that can be measured directly in 
spatially resolved images or indirectly through the modeling of the SED 
of the debris disk. Indeed, recent $\it{Spitzer}$ observations suggest that 
debris disks and giant planets co-exist and that inner gaps appear to be 
common in cold Kuiper Belt-like disks (Beichman et al.~\citeyear{beic05},
Kim et al.~\citeyear{kim05}). To aid in the interpretation of these 
observations, we have studied the efficiency of particle ejection 
and the resulting dust density contrast inside and outside the
orbit of the planet, as a function of the planet's mass and orbital 
elements and the particle size. 

\begin{center} {\it Acknowledgments} \end{center}
We thank Hal Levison for providing the SKEEL computer code, Alberto Noriega-Crespo, 
Dana Backman, George Rieke, Re'em Sari and Mark Sykes for useful discussion, and 
the anonymous referee for very helpful comments on how to improve the manuscript. 
This work is part of the {\it Spitzer} FEPS Legacy project (http://feps.as.arizona.edu). 
We acknowledge NASA for research support (contract 1224768 administered by JPL and 
grants NAG5-10343 and NAG5-11661), and IPAC, the $\it{Spitzer}$ Science Center and 
the Max-Plank-Institute in Heidelberg for providing access to their facilities. 
AMM is also supported by the Lyman Spitzer Fellowship at Princeton University.

\clearpage

\begin{deluxetable}{lcccccccc}
\tablewidth{0pc}
\tablehead{
\colhead{M$_{pl}$} & 
\colhead{$\it{a}$} & 
\colhead{$\it{e}$} & 
\colhead{$\beta$}  & 
\colhead{n$_{in}$} &
\colhead{n$_{col}$} & 
\colhead{n$_{1000}$(n$_{ejec}$)} &
\colhead{$\it\bar{v}_\infty$($\sigma_{\bar{v}_\infty})$} & 
\colhead{($\langle$$\it{v}$$_z^2\rangle$/$\langle$$\it{v}$$_{xy}^2\rangle$)$^{1/2}$}\\
\colhead{(M$_{Jup}$)} & 
\colhead{} & 
\colhead{} & 
\colhead{} & 
\colhead{(km/s)} & 
\colhead{} & 
\colhead{} & 
\colhead{} & 
\colhead{}}
\startdata
Solar & System & &	0.00156 &	1  &	4  & 95(29) & 2.3(1.8) & 0.10\\
1  &	1  &	0 &		&	6  &	5  & 89(72) & 4.8(4.5) & 0.20\\
3  &	1  &	0 &		&	19 &	1  & 80(62) & 3.4(3.5) & 0.002\\
10 & 	1  &	0 &		&	5  &    0  & 95(80) & 4.3(5.0) & 0.33\\
1  & 	5  & 	0 & 		& 	15 &	0  & 85(33) & 2.9(1.9) & 0.15\\
3  &	5  &	0 &		&	10 &	0  & 90(44) & 3.3(2.7) & 0.35\\
10 &	5  &	0 &		&	4  &	0  & 96(50) & 1.5(0.8) & 0.36\\
1  &	30 &	0 &		&	1  &    0  & 69     & \nodata  &\nodata\\
3  &	30 &    0 &		&	0  &    1  & 69     & \nodata  &\nodata\\
10 & 	30 &	0 &		&	0  &    2  & 69     & \nodata  &\nodata\\
Solar & System & &	0.00312 &	4  &	2  & 94(26) & 1.7(1.0) & 0.13\\
1  &	1  &	0 &		&	10 &	5  & 85(80) & 5.5(3.9) & 0.10\\
3  &	1  &	0 &		&	8  &	1  & 91(60) & 4.3(3.7) & 0.40\\
10 &	1  &	0 &		&	1  &	1  & 98(83) & 4.2(4.7) & 0.37\\
1  &	5  &	0 &		&	12 &	0  & 88(49) & 3.2(2.7) & 0.24\\
3  &	5  &	0 &		&	17 &	1  & 82(40) & 3.5(2.8) & 0.31\\
10 &	5  &	0 &		&	2  &	0  & 98(46) & 1.7(1.8) & 0.44\\
1   &	30   &	0  &		&	2  &	0  & 68     & \nodata  &\nodata\\
3   &	30   &	0  &		&	2  &	0  & 68     & \nodata  &\nodata\\
10   &	30   &	0  &		&	0  &	2  & 68     & \nodata  &\nodata\\
Solar &	System & &	0.00625 &	5  &	6  & 89(33) & 2.2(1.8) & 0.21\\
1  &	1  &	0 &		&	9  &	8  & 83(74) & 5.8(4.1) & 0.10\\
3  &	1  &	0 &		&	14 &	6  & 80(68) & 5.6(4.3) & 0.21\\
10 &	1  &	0 &		&	7  &	0  & 93(83) & 4.1(3.9) & 0.43\\
1  &	5  &	0 &		&	19 &	1  & 80(42) & 3.0(2.4) & 0.26\\
3  &	5  &	0 &		&	12 &	0  & 88(43) & 3.3(3.4) & 0.32\\
10 &	5  &	0 &		&	7  &	0  & 93(35) & 2.3(2.7) & 0.34\\
1   &	30   &	0  &		&	8  &	0  & 62     & \nodata  &\nodata\\
3   &	30   &	0  &		&	0  &	1  & 69     & \nodata  &\nodata\\
10   &	30   &	0  &		&	0  &	3  & 67     & \nodata  &\nodata\\
Solar &	System &  &	0.0125  &	8  &	3  & 89(32) & 2.2(1.9) & 0.08\\
1  &	1  &	0 &		&	17 &	2  & 81(74) & 5.0(3.7) & 0.08\\
3  &	1  &	0 &		&	5  &	8  & 87(82) & 5.9(3.1) & 0.15\\
10 &	1  &	0 &		&	5  &	0  & 95(81) & 3.8(3.7) & 0.38\\
1  &	5  &	0 &		&	14 &	0  & 86(63) & 2.5(1.5) & 0.10\\
3  &	5  &	0 &		&	8  &	0  & 92(39) & 2.7(2.0) & 0.36\\
10 &	5  &	0 &		&	3  &	0  & 97(48) & 2.0(2.1) & 0.43\\
1   &	30   &	0  &		&	4  &	0  & 66     & \nodata  &\nodata\\
3   &	30   &	0  &		&	5  &	0  & 65     & \nodata  &\nodata\\
10   &	30   &	0  &		&	1  &	2  & 67     & \nodata  &\nodata\\
Solar &	System &  &	0.025   &	15 &	1  & 84(31) & 2.0(1.7) & 0.09\\
1  &	1  &	0 &		&	23 &	1  & 76(65) & 5.9(4.2) & 0.08\\
3  &	1  &	0 &		&	7  &	10 & 83(79) & 7.0(3.2) & 0.13\\
10 &	1  &	0 &		&	7  &	0  & 93(81) & 3.7(3.4) & 0.39\\
1  &	5  &	0 &		&	17 &	0  & 83(56) & 2.5(1.9) & 0.14\\
3  &	5  &	0 &		&	14 &	3  & 83(42) & 2.7(2.0) & 0.35\\
10 &	5  &	0 &		&	9  &	0  & 91(55) & 2.4(2.4) & 0.45\\
1   &	30   &	0  &		&	11 &	0  & 59     & \nodata  &\nodata\\
3   &	30   &	0  &		&	2  &	1  & 67     & \nodata  &\nodata\\
10   &	30   &	0  &	        &	1  &	4  & 65     & \nodata  &\nodata\\
Solar & System &  &	0.044   &	19 &	3  & 78(28) & 2.1(1.6) & 0.08\\
1  &	1  &	0 &		&	39 &	3  & 58(53) & 5.8(4.7) & 0.10\\
3  &	1  &	0 &		&	7  &	2  & 91(89) & 6.5(3.8) & 0.06\\
10 &	1  &	0 &		&	11 &	0  & 89(78) & 4.6(5.6) & 0.31\\
1  &	5  &	0 &		&	25 &	0  & 75(53) & 2.6(1.6) & 0.09\\
3  &	5  &	0 &		&	12 &	0  & 88(52) & 3.7(3.0) & 0.27\\
10 &	5  &	0 &		&	5  &	0  & 95(48) & 2.0(2.2) & 0.41\\
1   &	30   &	0  &		&	11 &	1  & 58     & \nodata  &\nodata\\
3   &	30   &	0  &		&	0  &	3  & 67     & \nodata  &\nodata\\
10   &	30   &	0  &		&	2  &	1  & 67     & \nodata  &\nodata\\
Solar &	System &  &	0.1     &	21 &	3  & 76(48) & 2.0(1.5) & 0.05\\
1  &	1  &	0 &		&	43 &	2  & 55(53) & 7.0(4.8) & 0.10\\
3  &	1  &	0 &		&	18 &	5  & 77(75) & 7.9(4.1) & 0.06\\
10 &	1  &	0 &		&	5  &	1  & 94(90) & 5.7(4.1) & 0.14\\
1  &	5  &	0 &		&	38 &	1  & 61(37) & 3.0(2.0) & 0.11\\
3  &	5  &	0 &		&	14 &	2  & 84(68) & 2.9(1.7) & 0.16\\
10 &	5  &	0 &		&	10 &	0  & 90(55) & 2.2(1.6) & 0.35\\
1   &	30   &	0  &	  	&	11 &	2  & 57     & \nodata  &\nodata\\
3   &	30   &	0  &		&	3  & 	0  & 67     & \nodata  &\nodata\\
10   &	30   &	0  &		&	2  &	2  & 66     & \nodata  &\nodata\\
Solar &	System &  &	0.2     &	15 &	0  & 85(42) & 2.4(1.7) & 0.1\\
1  &	1  &	0 &		&	47 &	3  & 50(48) & 6.8(4.1) & 0.07\\
3  &	1  &	0 &		&	17 &	3  & 80(79) & 9.8(4.7) & 0.06\\
10 &	1  &	0 &		&	0  &	9  & 91(89) & 9.1(4.0) & 0.02\\
1  &	5  &	0 &		&	32 &	1  & 67(52) & 3.1(2.0) & 0.11\\
3  &	5  &	0 &		&	8  &	1  & 91(79) & 3.5(1.8) & 0.07\\
10 &	5  &	0 &		&	0  &	5  & 95(81) & 4.1(2.1) & 0.04\\
1   &	30   &	0  &		&	3  &	1  & 66     & \nodata  &\nodata\\
3   &	30   &	0  &		&	2  &	0  & 68     & \nodata  &\nodata\\
10   &	30   &	0  &		&	0  &	0  & 70     & \nodata  &\nodata\\
Solar & System &  &	0.4	&	11 &	0  & 89(58) & 3.3(2.1) & 0.10\\
1  &	1  &	0 &		&	39 &	5  & 56(53) & 11.5(6.8)& 0.13\\
3  &	1  &	0 &		&	7  &	5  & 88(86) & 11.0(6.9)& 0.10\\
10 &	1  &	0 &		&	0  &	7  & 93(92) & 12.2(6.5)& 0.03\\
1  &	5  &	0 &		&	24 &	1  & 75(67) & 4.0(2.6) & 0.13\\
3  &	5  &	0 &		&	5  &	0  & 95(90) & 4.6(2.5) & 0.07\\
10 &	5  &	0 &		&	0  &	5  & 95(90) & 5.6(2.9) & 0.05\\
1   &	30   &	0  &		&	0  &	0  & 70     & \nodata  &\nodata\\
3   &	30   &	0  &		&	0  &	0  & 70     & \nodata  &\nodata\\
10   &	30   &	0  &		&	0  &    0  & 70     & \nodata  &\nodata\\
\tablenotetext{~}
{The first three columns list the parameters of the planetary system:
$M_{pl}$ is the mass of the planet in Jupiter-masses,
$a$ is the planet's semimajor axis and $e$ its orbital eccentricity;
rows labeled "Solar System" are for models which include the seven
major planets of the Solar system (from Venus to Neptune) with masses 
and orbital parameters from the Astronomical Almanac 2000. For all models, 
the central star is assumed to be solar-type. 
The fourth column lists the dust particle's $\beta$ value,
and the remaining columns list the final fates of the dust
particles in each model.
n$_{in}$ is the number of particles that drift all the way to the
inner cut-off distance (0.1 AU from the central star for the models
with the planet at 1 AU, and 0.5 AU for the rest);
n$_{col}$ is number of particles that collide with the planet(s);
n$_{1000}$ is number of particles that reach 1000 AU;
n$_{ejec}$ is number of particles on hyperbolic orbits (E$>$0);
$\it\bar{v}_\infty$ is the mean value of the velocity at infinity,
(2E)$^{1/2}$, of the particles on hyperbolic orbits, and
$\sigma_{\bar{v}_\infty}$ is its standard deviation.
In the Solar system models, initial conditions of the dust particles
are derived from assumed parent bodies having a distribution similar
to the Solar system's KBOs, with $\it{a}$ in the range 35--50 AU,
$\it{e}$ such that perihelion distance is in the range 35--50 AU and
$\it{i}$ in the range 0--17$^\circ$.
For all single-planet models, the parent bodies were assumed distributed
with $\it{a}$ in the range 35--50 AU, $\it{e}$ in the range 0--0.05
and $\it{i}$ in the range 0--0.05 radians.
In each model we simulated 100 dust particles, with the exception of
the single-planet models with the planet at 30 AU; in the latter models,
we simulated 70 particles from parent bodies assumed to have $a$ in the
range 40--50 AU, since closer objects would be destabilized by the
planet's perturbations.}
\enddata
\end{deluxetable}

\clearpage

\begin{deluxetable}{lcccccccc}
\tablewidth{0pc}
\tablehead{
\colhead{M$_{pl}$} & 
\colhead{$\it{a}$} & 
\colhead{$\it{e}$} & 
\colhead{$\beta$}  & 
\colhead{n$_{in}$} &
\colhead{n$_{col}$} & 
\colhead{n$_{1000}$(n$_{ejec}$)} &
\colhead{$\it\bar{v}_\infty$($\sigma_{\bar{v}_\infty})$} & 
\colhead{($\langle$$\it{v}$$_z^2\rangle$/$\langle$$\it{v}$$_{xy}^2\rangle$)$^{1/2}$}\\
\colhead{(M$_{Jup}$)} & 
\colhead{} & 
\colhead{} & 
\colhead{} & 
\colhead{(km/s)} & 
\colhead{} & 
\colhead{} & 
\colhead{} & 
\colhead{}}
\startdata
0.03 &	1 &	0 &	0.044 &		100 &	0  & 0(0)    & \nodata  &\nodata\\
0.1  &	  &	  &	      &		98  &	2  & 0(0)    & \nodata	&\nodata\\
0.3  &	  &	  &	      &		89  &	4  & 7(4)    & 5.2(3.7) & 0.11\\
1    &	  &	  &	      &		39  &	3  & 58(53)  & 5.8(4.7) & 0.10\\
3    &	  &	  &	      &		7   &	2  & 91(89)  & 6.4(3.8) & 0.06\\
10   &	  &	  &	      &		11  &	0  & 89(78)  & 4.6(5.6) & 0.31\\
0.03 &5.2 &	0 &	0.044 &		100 &	0  & 0(0)    & \nodata	&\nodata\\
0.1  &	  &	  &	      &		100 &	0  & 0(0)    & \nodata  &\nodata\\
0.3  &	  &	  &	      &		78  &	0  & 22(11)  & 2.9(2.0) & 0.05\\
1    &	  &	  &	      &		25  &	0  & 75(53)  & 2.6(1.6) & 0.09\\
3    &	  &	  &	      &		12   &	0  & 88(52)  & 3.7(3.0) & 0.27\\
10   &	  &	  &	      &		5   &	0  & 95(48)  & 2.0(2.2) & 0.41\\
0.03 &	10&	0 &	0.044 &		100 &	0  & 0(0)    & \nodata	&\nodata\\
0.1  &	  &	  &	      &		95  &	1  & 4(1)    & 1.1      & 1.36\\
0.3  &	  &	  &	      &		71  &	0  & 29(11)  & 2.0(1.3) & 0.10\\
1    &	  &	  &	      &		18  &	1  & 81(40)  & 1.8(1.1) & 0.08\\
3    &	  &	  &	      &		12  &	0  & 88(56)  & 2.0(1.1) & 0.14\\
10   &	  &	  &	      &		1   &	0  & 99(44)  & 1.5(0.9) & 0.16\\
0.03 &	20&	0 &	0.044 &		99  &	1  & 0(0)    & \nodata	&\nodata\\
0.1  &	  &	  &	      &		98  &	0  & 2(1)    & 3.4	& 0.00\\
0.3  &	  &	  &	      &		56  &	2  & 42(11)  & 1.7(1.1) & 0.06\\
1    &	  &	  &	      &		20  &	0  & 80(28)  & 1.2(0.7) & 0.13\\
3    &	  &	  &	      &		6   &	1  & 93(40)  & 1.3(0.8) & 0.10\\
10   &	  &	  &	      &		0   &	0  & 100(35) & 1.1(0.6) & 0.12\\
0.03  &	30&	0 &	0.044 &		70  &	0  & 0       & \nodata	&\nodata\\
0.1   &	  &       &	      &	        68  &	0  & 2       & \nodata	&\nodata\\
0.3   &	  &	  &	      &		41  &	0  & 29      & \nodata	&\nodata\\
1     &	  & 	  &	      &		11  &	1  & 58      & \nodata	&\nodata\\
3     &	  &	  &	      &		0   &	3  & 67      & \nodata	&\nodata\\
10    &   &       &	      &		2   &   1  & 67      & \nodata	&\nodata\\
\tablenotetext{~}
{The column headings are the same as Table 1, but in these models
the parent bodies of the dust particles are assumed to have an orbital
distribution similar to the solar system's KBOs, with $a$ in the range
35--50 AU, $e$ such that periastron distance is in the range 35--50 AU,
and $i$ in the range 0--17$^\circ$.}
\enddata
\end{deluxetable}

\clearpage

\begin{deluxetable}{lcccccccc}
\tablewidth{0pc}
\tablehead{
\colhead{M$_{pl}$} & 
\colhead{$\it{a}$} & 
\colhead{$\it{e}$} & 
\colhead{$\beta$}  & 
\colhead{n$_{in}$} &
\colhead{n$_{col}$} & 
\colhead{n$_{1000}$(n$_{ejec}$)} &
\colhead{$\it\bar{v}_\infty$($\sigma_{\bar{v}_\infty})$} & 
\colhead{($\langle$$\it{v}$$_z^2\rangle$/$\langle$$\it{v}$$_{xy}^2\rangle$)$^{1/2}$}\\
\colhead{(M$_{Jup}$)} & 
\colhead{} & 
\colhead{} & 
\colhead{} & 
\colhead{(km/s)} & 
\colhead{} & 
\colhead{} & 
\colhead{} & 
\colhead{}}
\startdata
1    &  5 &	0   &	0.044 &		25  &	0  & 75(53)  & 2.6(1.6) & 0.09\\
     &	  &	0.1 &	      &		22  &	1  & 77(50)  & 2.4(1.7) & 0.11\\
     &	  &	0.2 &	      &		36  &	1  & 63(42)  & 2.6(1.6) & 0.15\\
     &	  &	0.3 &	      &		32  &	1  & 67(30)  & 2.7(1.7) & 0.10\\
     &	  &	0.4 &	      &		55  &	1  & 44(27)  & 2.8(1.9) & 0.11\\
     &	  &	0.5 &	      &		65  &	0  & 35(20)  & 3.1(1.9) & 0.10\\
1    &	10&	0   &	0.044 &		18  &	1  & 81(40)  & 1.8(1.1) & 0.08\\
     &	  &	0.1 &	      &		24  &	2  & 74(33)  & 1.6(1.1) & 0.09\\
     &	  &	0.2 &	      &		19  &	1  & 80(37)  & 1.7(1.2) & 0.09\\
     &	  &	0.3 &	      &		31  &	0  & 69(23)  & 1.7(1.3) & 0.12\\
     &	  &	0.4 &	      &		45  &	2  & 53(19)  & 2.2(1.2) & 0.10\\
     &	  &	0.5 &	      &		56  &	1  & 43(16)  & 1.8(0.9) & 0.15\\
1    &	20&	0   &	0.044 &		20  &	0  & 80(28)  & 1.2(0.8) & 0.13\\
     &	  &	0.1 &	      &		18  &	0  & 82(19)  & 1.7(1.7) & 0.16\\
     &	  &	0.2 &	      &		27  &	1  & 72(13)  & 1.5(0.9) & 0.16\\
     &	  &	0.3 &	      &		27  &	0  & 73(18)  & 1.6(1.3) & 0.16\\
     &	  &	0.4 &	      &		38  &	0  & 62(22)  & 2.2(2.2) & 0.16\\
     &	  &	0.5 &	      &		47  &	1  & 52(14)  & 1.8(1.3) & 0.22\\
\tablenotetext{~}
{The initial conditions of dust particles are the same as in Table 2 models.}
\enddata
\end{deluxetable}

\clearpage

\begin{deluxetable}{lccccc}
\tablewidth{0pc}
\tablehead{
\colhead{M$_{pl}$(M$_{Jup}$)} & 
\colhead{$\it{a}$} & 
\colhead{$\it{e}$} & 
\colhead{$\beta$}  & 
\colhead{r$_{gap}$ (AU)} &
\colhead{r$_{gap}$/$\it{a}$}}
\startdata
1  &	1  &	0 & 0.00156 & 0.8  & 0.8\\
3  &	1  &	0 &         & 0.8  & 0.8\\
10 & 	1  &	0 &         & 1.2  & 1.2\\
1  & 	5  & 	0 &         & 4.2  & 0.8\\
3  &	5  &	0 &         & 4.8  & 0.9\\
10 &	5  &	0 &         & 6.4  & 1.2\\
1  &	30 &	0 &         & 25.2 & 0.8\\
3  &	30 &    0 &         & 25.2 & 0.8\\
10 & 	30 &	0 &         & 32.2 & 1.1\\
1  &	1  &	0 & 0.00312 & 0.8  & 0.8\\ 
3  &	1  &	0 &         & 0.8  & 0.8\\
10 & 	1  &	0 &         & 1.2  & 1.2\\
1  & 	5  & 	0 &         & 4.4  & 0.8\\
3  &	5  &	0 &         & 4.6  & 0.9\\
10 &	5  &	0 &         & 6.4  & 1.2\\
1  &	30 &	0 &         & 25.2 & 0.8\\
3  &	30 &    0 &         & 25.2 & 0.8\\
10 & 	30 &	0 &         & 35.5 & 1.2\\
1  &	1  &	0 & 0.00625 & 0.8  & 0.8\\ 
3  &	1  &	0 &         & 0.8  & 0.8\\
10 & 	1  &	0 &         & 1.2  & 1.2\\
1  & 	5  & 	0 &         & 4.4  & 0.8\\
3  &	5  &	0 &         & 4.6  & 0.9\\
10 &	5  &	0 &         & 6.1  & 1.2\\
1  &	30 &	0 &         & 25.2 & 0.8\\
3  &	30 &    0 &         & 25.2 & 0.8\\
10 & 	30 &	0 &         & 37.2 & 1.2\\
1  &	1  &	0 & 0.0125  & 0.8  & 0.8\\
3  &	1  &	0 &         & 0.8  & 0.8\\
10 & 	1  &	0 &         & 1.1  & 1.1\\
1  & 	5  & 	0 &         & 4.4  & 0.8\\
3  &	5  &	0 &         & 4.1  & 0.8\\
10 &	5  &	0 &         & 6.1  & 1.2\\
1  &	30 &	0 &         & 25.2 & 0.8\\
3  &	30 &    0 &         & 25.2 & 0.8\\
10 & 	30 &	0 &         & 37.2 & 1.2\\
1  &	1  &	0 & 0.025   & 0.8  & 0.8\\
3  &	1  &	0 &         & 0.8  & 0.8\\
10 & 	1  &	0 &         & 1.2  & 1.2\\
1  & 	5  & 	0 &         & 4.4  & 0.8\\
3  &	5  &	0 &         & 4.4  & 0.8\\
10 &	5  &	0 &         & 6.4  & 1.2\\
1  &	30 &	0 &         & 24.0 & 0.8\\
3  &	30 &    0 &         & 25.2 & 0.8\\
10 & 	30 &	0 &         & 37.2 & 1.2\\
1  &	1  &	0 & 0.044   & 0.8  & 0.8\\
3  &	1  &	0 &         & 0.8  & 0.8\\
10 & 	1  &	0 &         & 1.4  & 1.4\\
1  & 	5  & 	0 &         & 4.2  & 0.8\\
3  &	5  &	0 &         & 4.2  & 0.8\\
10 &	5  &	0 &         & 5.8  & 1.1\\
1  &	30 &	0 &         & 24.0 & 0.8\\
3  &	30 &    0 &         & 25.2 & 0.8\\
10 & 	30 &	0 &         & 39.1 & 1.3\\
1  &	1  &	0 & 0.1     & 0.8  & 0.8\\
3  &	1  &	0 &         & 0.8  & 0.8\\
10 & 	1  &	0 &         & 1.2  & 1.2\\
1  & 	5  & 	0 &         & 4.1  & 0.8\\
3  &	5  &	0 &         & 4.1  & 0.8\\
10 &	5  &	0 &         & 6.1  & 1.2\\
1  &	30 &	0 &         & 24.0 & 0.8\\
3  &	30 &    0 &         & 25.2 & 0.8\\
10 & 	30 &	0 &         & 35.5 & 1.2\\
1  &	1  &	0 & 0.2     & 0.8  & 0.8\\
3  &	1  &	0 &         & 0.8  & 0.8\\
10 & 	1  &	0 &         & 1.2  & 1.2\\
1  & 	5  & 	0 &         & 4.0  & 0.8\\
3  &	5  &	0 &         & 4.0  & 0.8\\
10 &	5  &	0 &         & 6.8  & 1.3\\
1  &	30 &	0 &         & 22.9 & 0.8\\
3  &	30 &    0 &         & 24.0 & 0.8\\
10 & 	30 &	0 &         & 37.2 & 1.2\\
1  &	1  &	0 & 0.4     & 0.7  & 0.7\\
3  &	1  &	0 &         & 0.7  & 0.7\\
10 & 	1  &	0 &         & 2.1  & 2.1\\
1  & 	5  & 	0 &         & 3.8  & 0.7\\
3  &	5  &	0 &         & 3.6  & 0.7\\
10 &	5  &	0 &         & 6.4  & 1.2\\
1  &	30 &	0 &         & 24.0 & 0.8\\
3  &	30 &    0 &         & 39.1 & 1.3\\
10 & 	30 &	0 &         & 37.2 & 1.2\\
\tablenotetext{~}
{r$_{gap}$ is the astrocentric distance of the gap in AU, determined by the radius at 
which the surface density from the numerical results decreases by more than 90\%.}
\enddata
\end{deluxetable}

\clearpage

\begin{figure}
\plotone{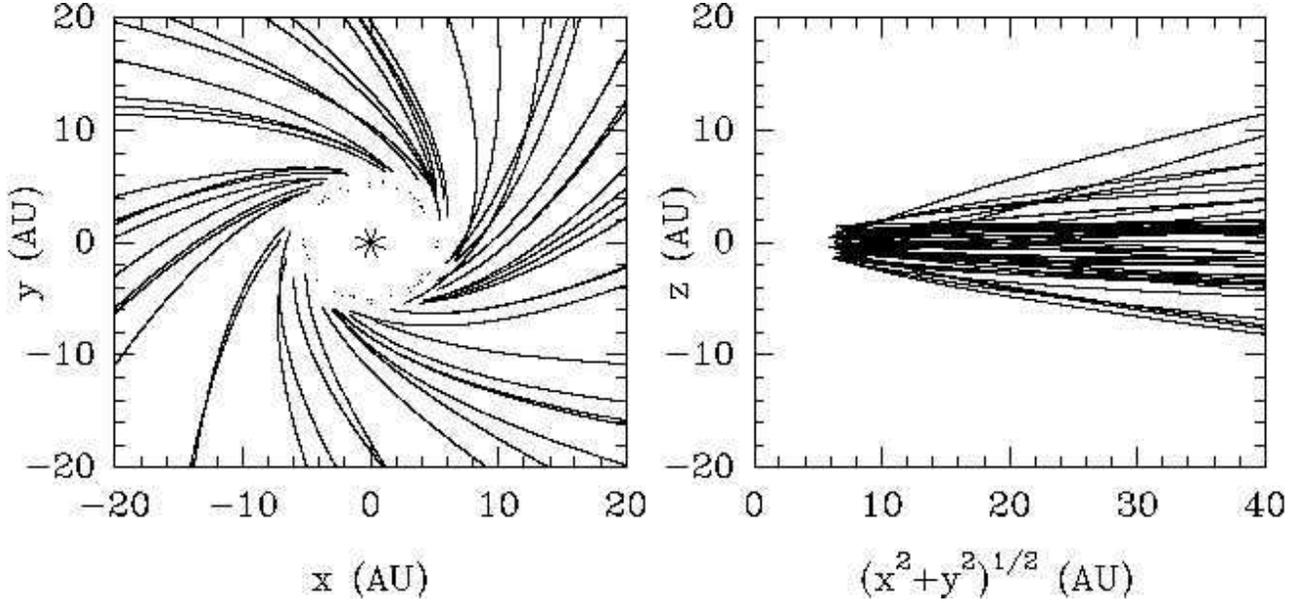}
\caption{Trajectories of the particles that reach 1000 AU after scattering by
Jupiter. These particles have $\beta$=0.2 and their paths are shown 
just after the last encounter with the planet.
(left) in the XY plane;~the dots represent the position of Jupiter
at the time of last encounter;
~(right) in the RZ plane, where R=(x$^2$+y$^2$)$^{1/2}$ is the in-plane 
heliocentric distance and Z is the off-plane out-of-ecliptic distance.}
\end{figure}

\begin{figure}
\epsscale{1.0}
\plotone{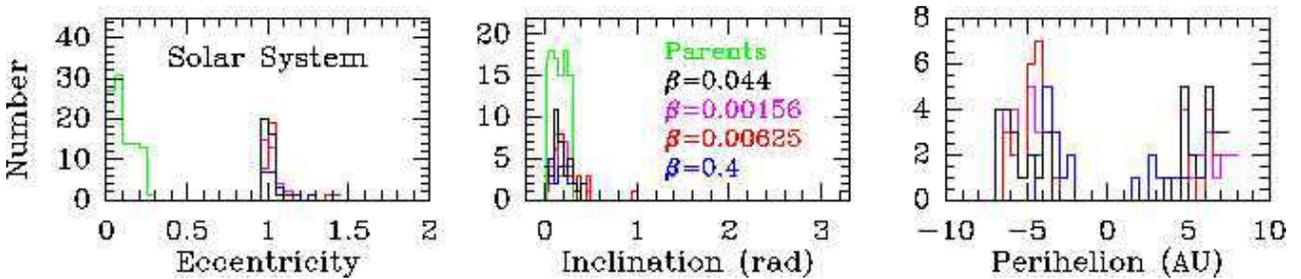}
\caption{Distribution of eccentricity (left), inclination 
(center) and perihelion (right) of the particles ejected by Jupiter in the solar
system models. Three different particles sizes are shown, corresponding to 
$\beta$-values of 0.044 (black), 0.00156 (magenta), 0.00625 (red) and 0.4 (blue).
The green lines show the distributions for the parent
bodies with $\it{a}$=35--50 AU, $\it{e}$ such that
perihelion=35--50 AU and $\it{i}$=0--17$^\circ$.}
\end{figure}

\clearpage

\begin{figure}
\epsscale{0.5}
\plotone{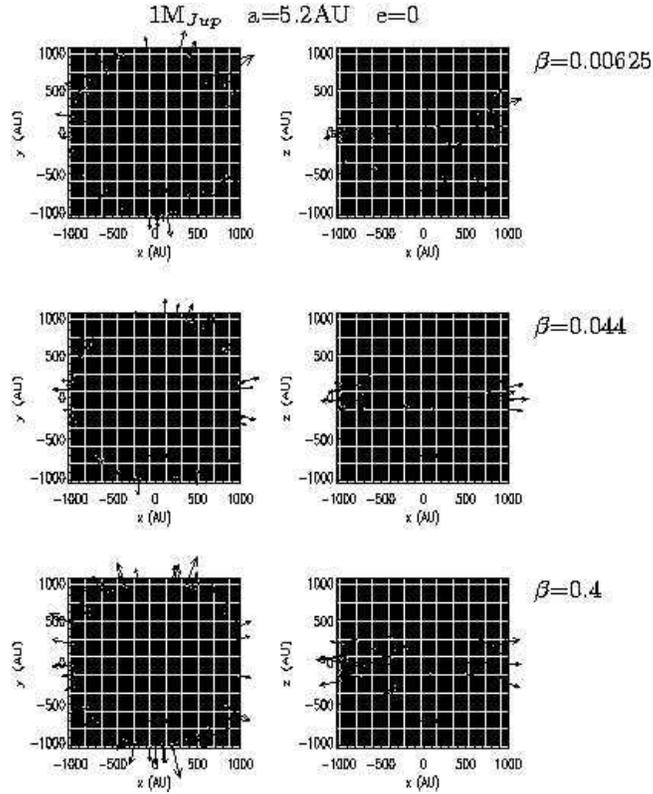}
\caption{Escaping dust particles of three different sizes (i.e. three different 
$\beta$-values), shown in the XY plane (left) and XZ plane (right). These are models
with a 1M$_{Jup}$ planet with $\it{a}$=5.2 AU and $\it{e}$=0. The magnitude of the 
particle velocity at infinity is indicated by the length of the arrows; the velocity 
scale of 10 km/s is indicated by the size of the large arrow at the bottom-center 
in each panel. In all cases, the dust-producing planetesimals are randomly distributed 
with $\it{a}$=35--50 AU, $\it{e}$=0--0.05 and $\it{i}$=0--0.05 radians.}
\end{figure}

\clearpage

\begin{figure}
\epsscale{0.5}
\plotone{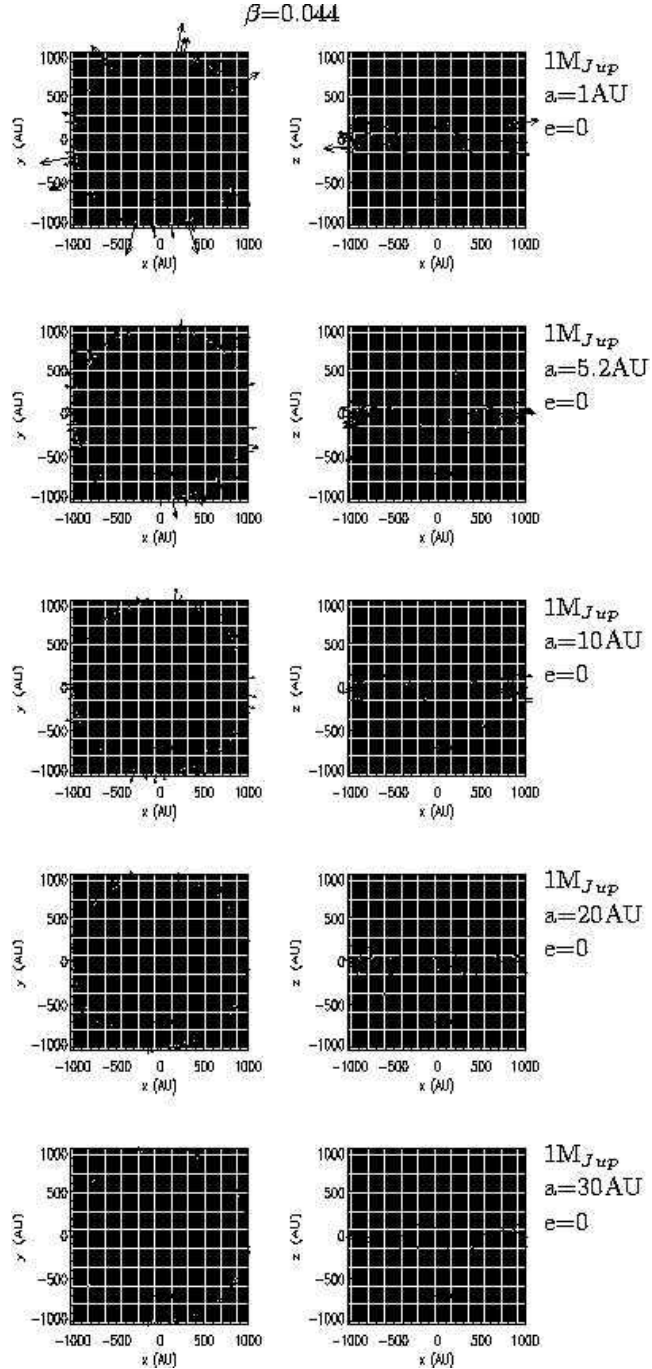}
\caption{Escaping dust particles 
(with $\beta$=0.044) in five different planetary systems, shown in the XY plane (left) 
and XZ plane (right). From top to bottom, the panels correspond to models with a 
1M$_{Jup}$ planet in a circular orbit and with semimajor axis of 1 AU, 5.2 AU, 
10 AU, 20 AU and 30 AU, respectively. The magnitude of the 
particle velocity at infinity is indicated by the length of the arrows; 
the velocity scale of 10 km/s is indicated by the size of the 
large arrow at the bottom-center in each panel. In all cases, the dust-producing 
planetesimals are randomly distributed with $\it{a}$=35--50AU, $\it{q}$=35--50 AU and 
$\it{i}$=0--17$^\circ$, similarly to the KBOs.}
\end{figure}

\clearpage

\begin{figure}
\epsscale{0.5}
\plotone{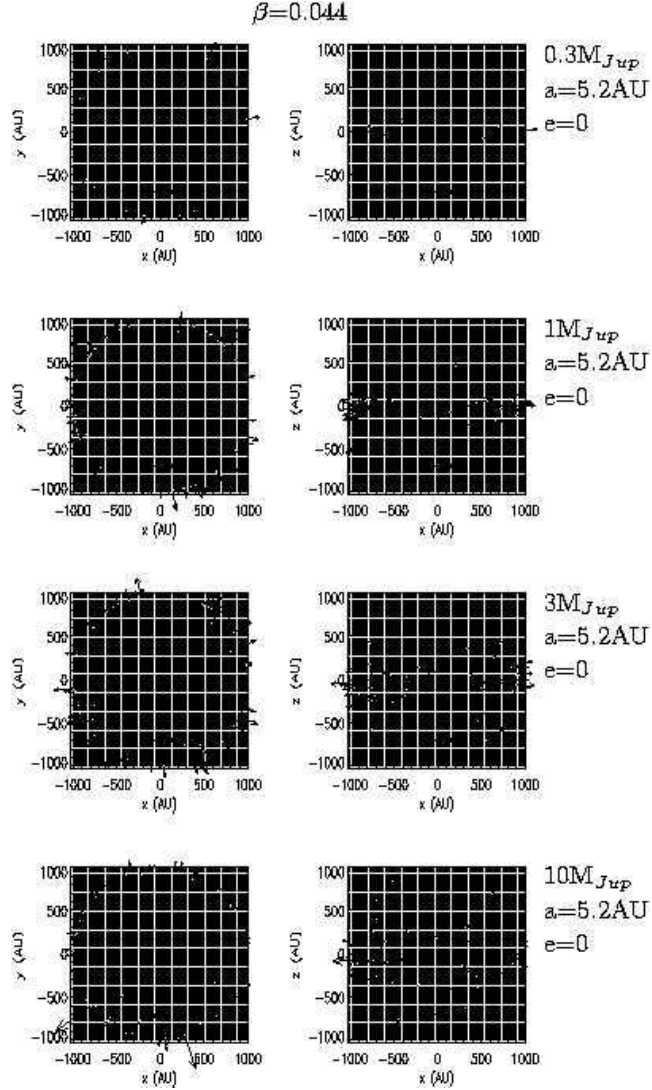}
\caption{Escaping dust particles 
(with $\beta$=0.044) in four different planetary systems, shown in the XY plane (left) 
and XZ plane (right). From top to bottom, the panels correspond to models with a 
single planet with $\it{a}$=5.2 AU and $\it{e}$=0 and a mass of
0.3M$_{Jup}$, 1M$_{Jup}$, 3M$_{Jup}$ and 10M$_{Jup}$, respectively. The magnitude of the 
particle velocity at infinity is indicated by the length of the arrows; 
the velocity scale of 10 km/s is indicated by the size of the 
large arrow at the bottom-center in each panel. In all cases, the dust-producing 
planetesimals are randomly distributed with $\it{a}$=35--50AU, $\it{q}$=35--50 AU and 
$\it{i}$=0--17$^\circ$, similarly to the KBOs.}
\end{figure}

\clearpage

\begin{figure}
\epsscale{0.5}
\plotone{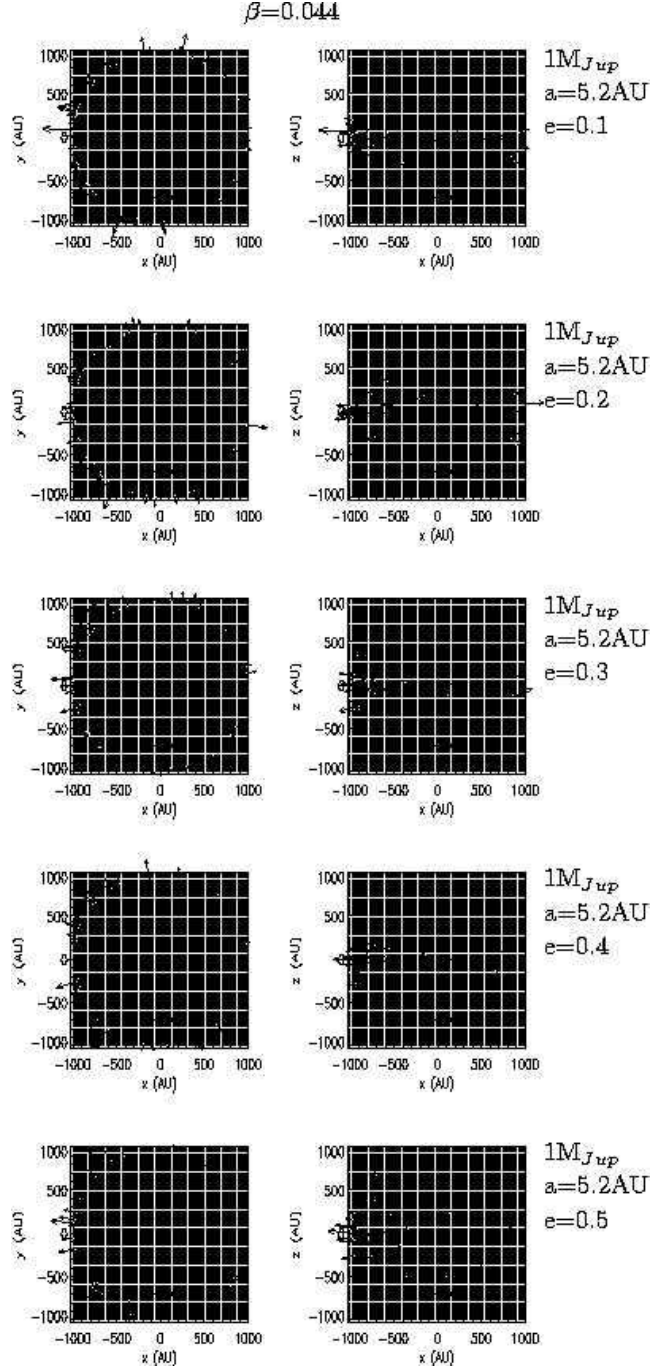}
\caption{Escaping dust particles 
(with $\beta$=0.044) in five different planetary systems, shown in the XY plane (left) 
and XZ plane (right). From top to bottom, the panels correspond to models with a 
single 1M$_{Jup}$ planet with $\it{a}$=5.2 AU and eccentricity of 
0.1, 0.2, 0.3, 0.4 and 0.5, respectively. The magnitude of the 
particle velocity at infinity is indicated by the length of the arrows; 
the velocity scale of 10 km/s is indicated by the size of the 
large arrow at the bottom-center in each panel. In all cases, the dust-producing 
planetesimals are randomly distributed with $\it{a}$=35--50AU, $\it{q}$=35--50 AU and 
$\it{i}$=0--17$^\circ$, similarly to the KBOs.}
\end{figure}

\clearpage

\begin{figure}
\epsscale{0.75}
\plotone{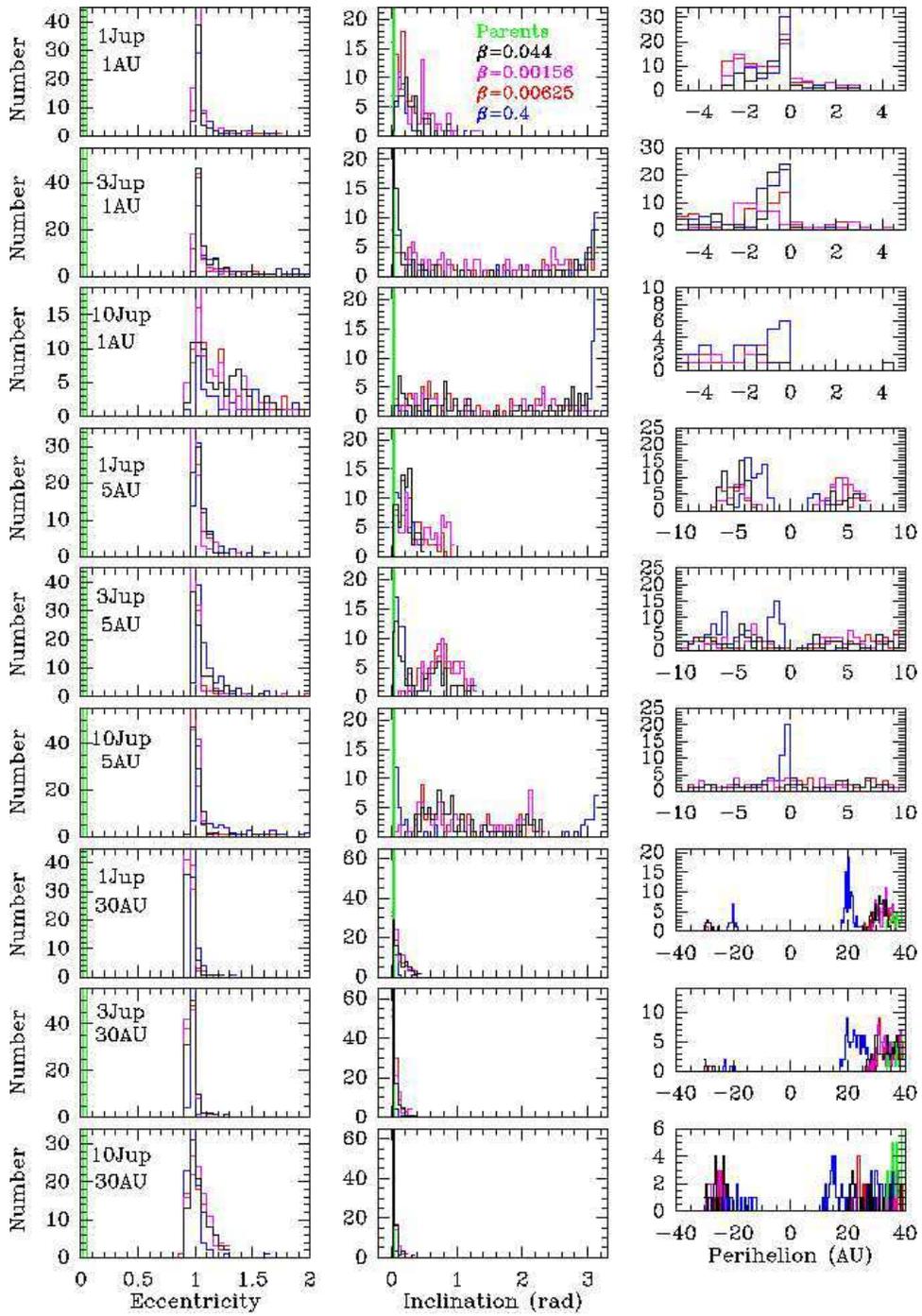}
\caption{Distribution of eccentricity (left), inclination 
(center) and perihelion (right) of the ejected particles that reach 1000 AU, for
three different particles sizes, corresponding to $\beta$-values of 0.044 (black), 
0.00156 (magenta), 0.00625 (red) and 0.4 (blue). The green lines show the distributions for the parent
bodies with $\it{a}$=35--50 AU, $\it{e}$=0--0.05 and 
$\it{i}$=0--0.05 radians.}
\end{figure}

\clearpage

\begin{figure}
\epsscale{0.75}
\plotone{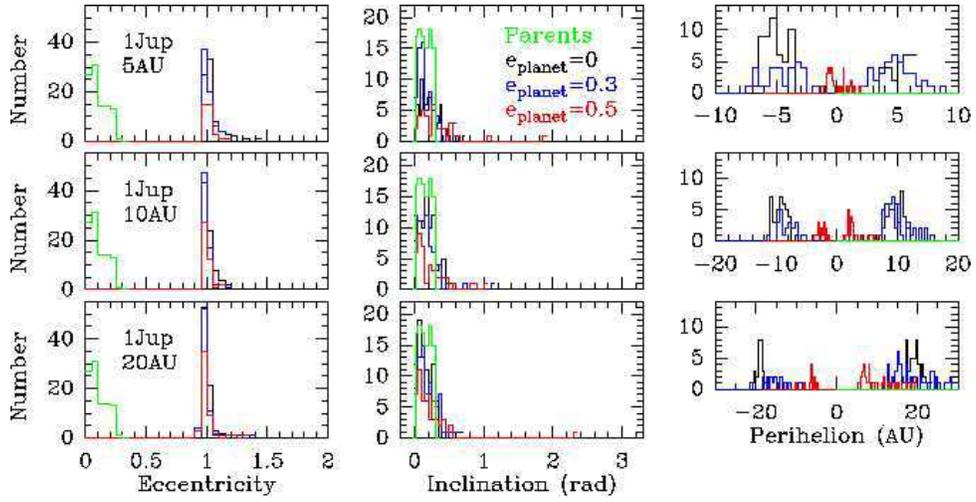}
\caption{Distribution of eccentricity (left), inclination 
(center) and perihelion (right) of the ejected particles that reach 1000 AU (with
$\beta$=0.044), for a system with a 1M$_{Jup}$ planet with eccentricity
of 0 (black), 0.3 (blue) and 0.5 (red). The green lines show the distributions for the parent
bodies with $\it{a}$=35--50 AU, $\it{e}$ such that
perihelion=35--50 AU and $\it{i}$=0--17$^\circ$.}
\end{figure}

\clearpage

\begin{figure}
\epsscale{1.}
\plotone{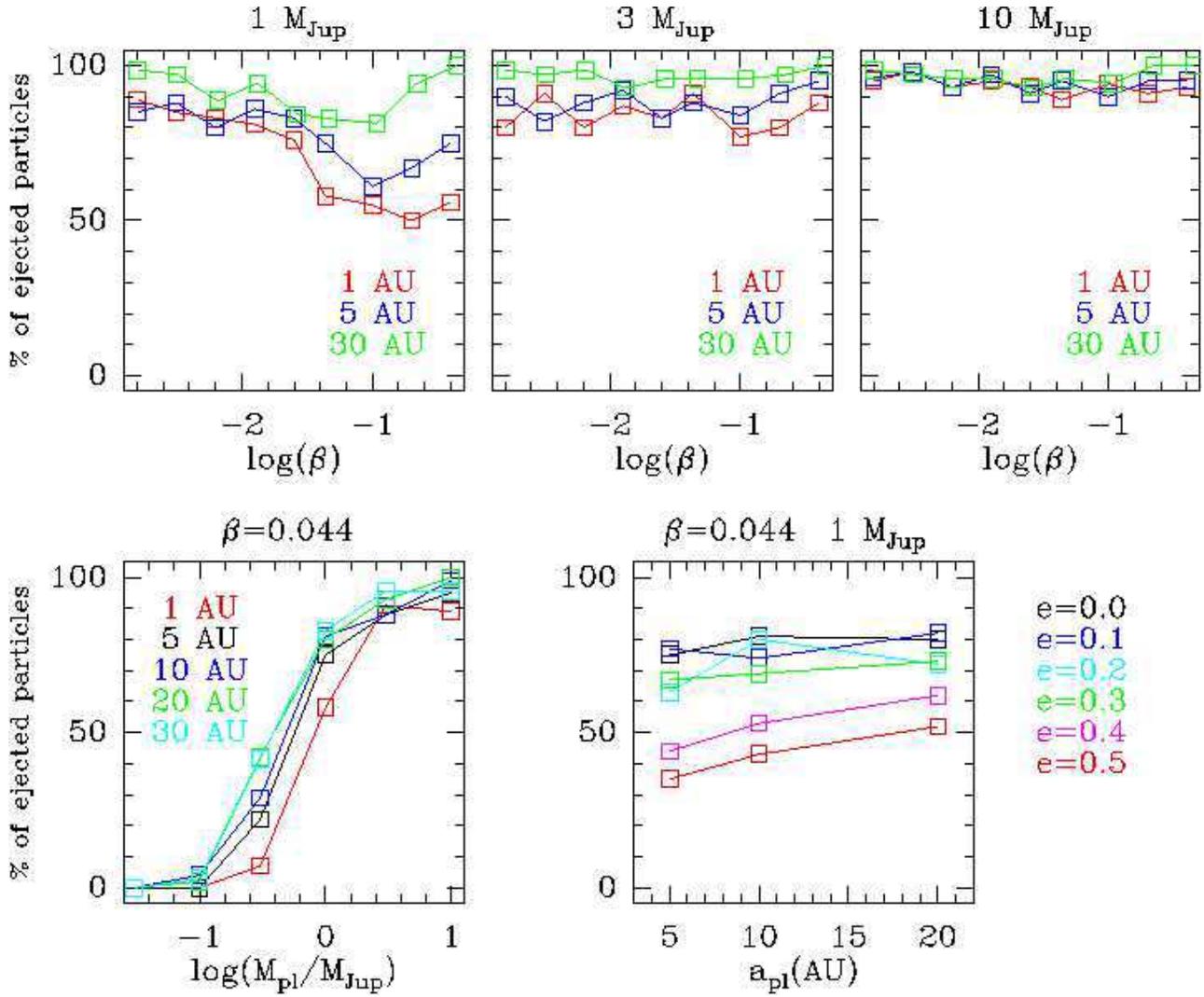}
\caption{Percentage of ejected particles (n$_{1000}$) as a function of planet's mass, planet's
semimajor axis and eccentricity and particle size. The models with the planet at 30 AU were 
based on 70 dust parent bodies between 40 and 50 AU.}
\end{figure}

\clearpage

\begin{figure}
\epsscale{1.}
\plotone{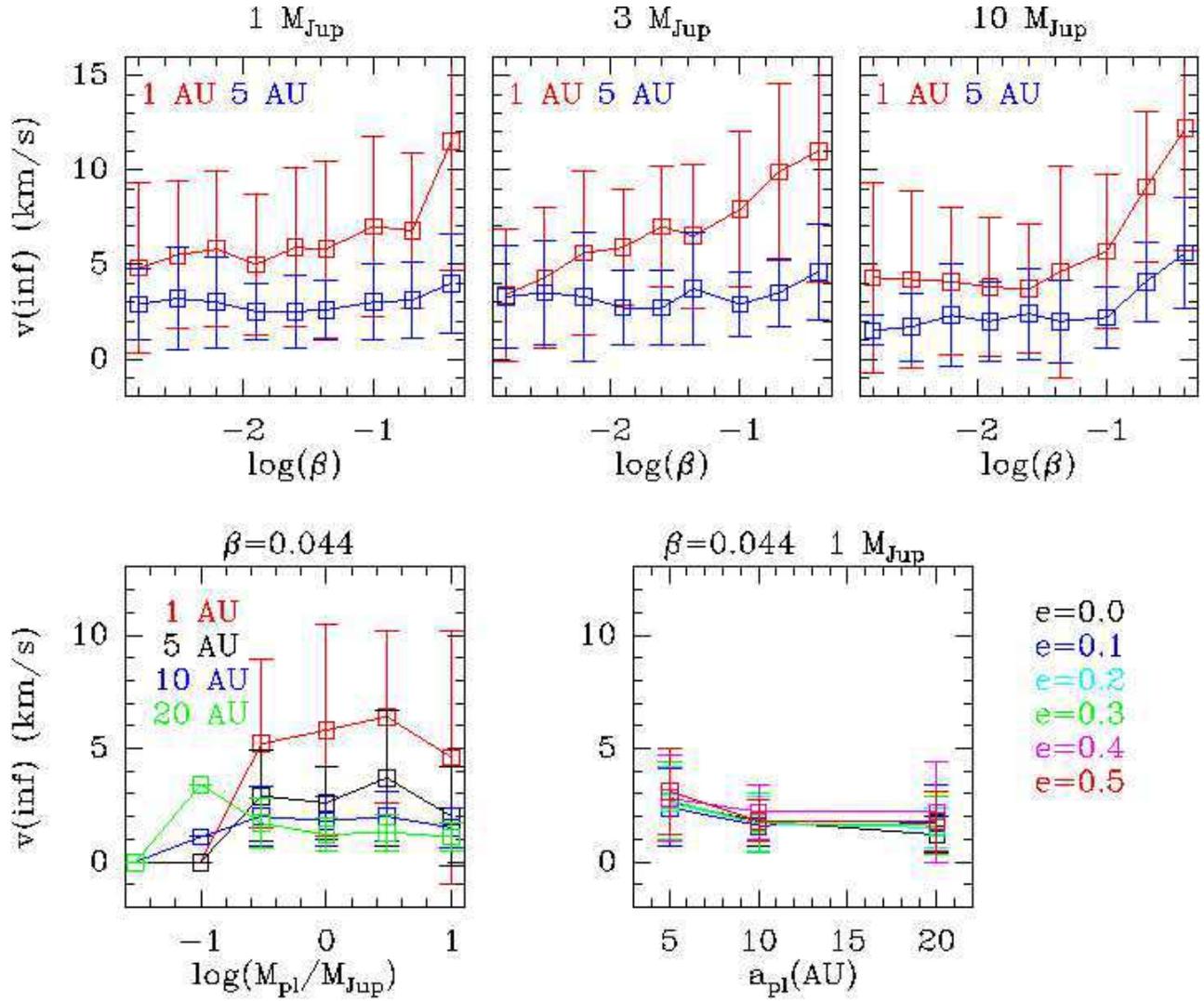}
\caption{Velocity at infinity of particles in hyperbolic orbits as a function of planet's 
mass, planet's semimajor axis and eccentricity and particle size.}
\end{figure}

\clearpage

\begin{figure}
\epsscale{1.}
\plotone{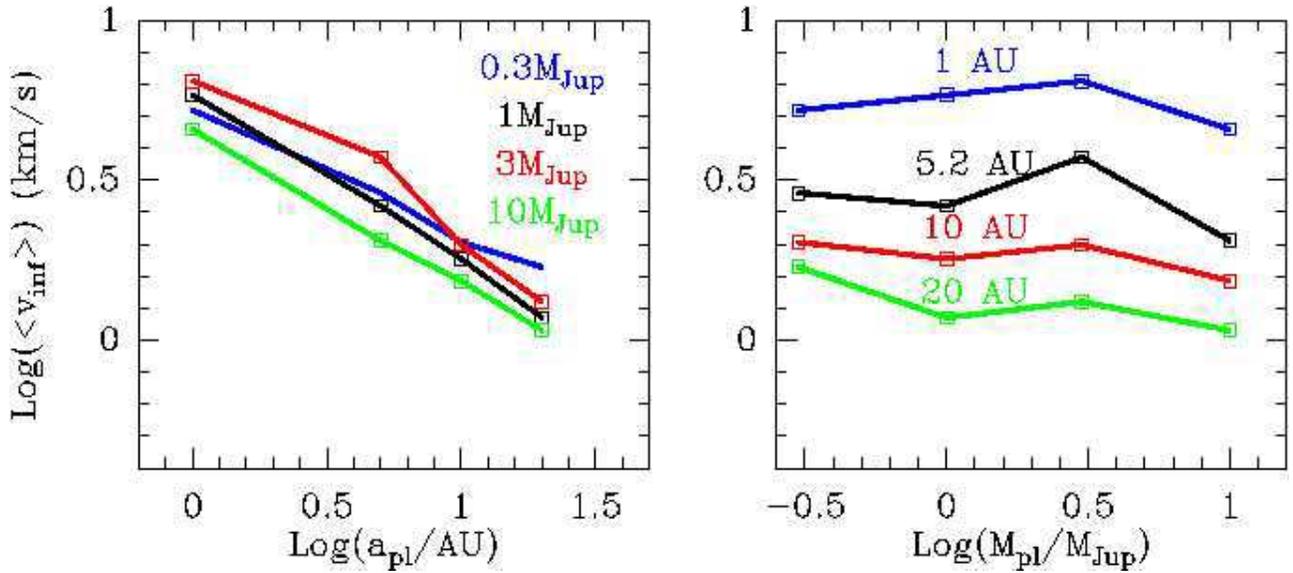}
\caption{Logarithmic plot of the average velocity at infinity of ejected 
particles of $\beta$=0.044 as a function of planet semimajor axis 
and planet mass. (left) The blue, black, red and green lines correspond to 
planet masses of 0.3M$_{Jup}$, 1M$_{Jup}$, 3M$_{Jup}$ and 10M$_{Jup}$, respectively.
(right) The blue, black, red, green and light blue lines correspond to 
models with a planet at 1AU, 5.2 AU, 10 AU and 20 AU, respectively.
The parent bodies of the dust particles are distributed like the KBOs,
with $\it{a}$=35--50 AU, $\it{e}$ such that
perihelion=35--50 AU and $\it{i}$=0--17$^\circ$.}
\end{figure}

\clearpage

\begin{figure}
\epsscale{0.75}
\plotone{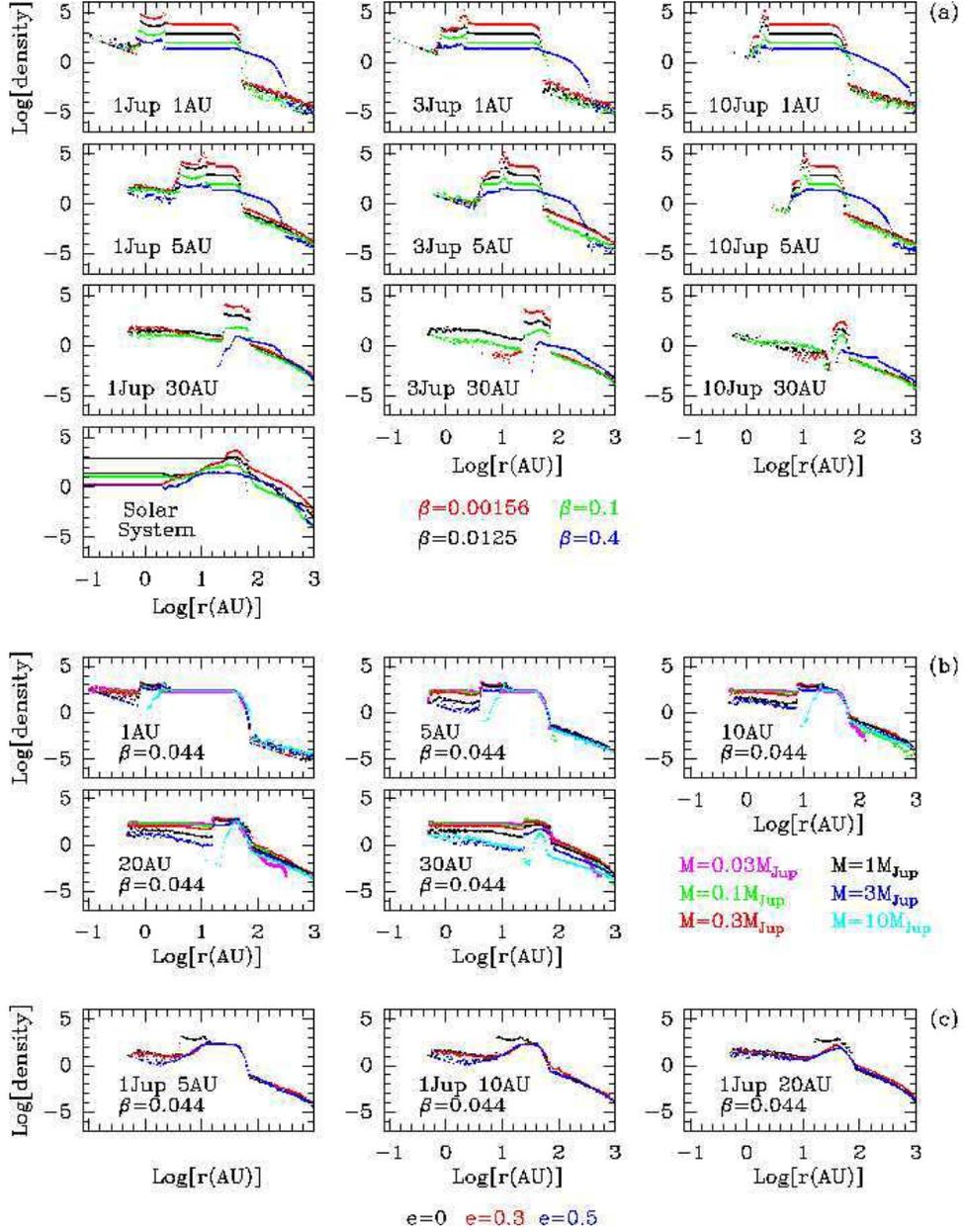}
\caption{(a: top 10 panels) Surface density distributions of dust particles with four 
different $\beta$ values (represented by different colors), for different planetary 
systems (indicated in the individual panels).
The units are number of particles per AU$^2$ for a dust production rate of 100 particles
per 1000 years (to be later scaled to the correct dust production rate or total disk mass).
(b: middle 5 panels) Surface density distributions of dust particles with $\beta$=0.044, for 
different planet masses (represented by different colors), located at five different
semimajor axes (indicated in the individual panels).
(c: bottom 4 panels) Surface density distributions of dust particles with $\beta$=0.044 and
a 1 M$_{Jup}$ planet, locatated at four different semimajor axes (indicated in the individual 
panels), and with three different eccentricities (represented by different colors).} 
\end{figure}

\end{document}